# Analytical form of current-voltage characteristic of a parallel-plane ionization chamber

**Dimitar G STOYANOV**

Faculty of Engineering and Pedagogy in Sliven, Technical University of Sofia
59, Bourgasko Shaussee Blvd, 8800 Sliven, BULGARIA

E-mail: dgstoyanov@abv.bg

**Abstract.** The elementary processes taking place in the formation of charged particles and their flow in the ionization chamber are considered. On the basic of particles and charges balance a differential equation describing the distribution of current densities in the ionization chamber volume is obtained. As a result of the differential equation solution an analytical form of the current-voltage characteristic of a parallel-plane ionization chamber is obtained.

## 1. Introduction

The ionization chamber is a simple and reliable device for measuring the characteristics of beams of particles and radiations. This refers to either beams of radioactive emission or X-rays, synchrotron radiation and beams of electrons and protons in different accelerators. The ionization chamber has been turned into an irreplaceable tool in such investigations because of which it is necessary the properties and the parameters of the ionization chamber to be got to know in different working regimes.

The ionization chamber represents a volume filled with gas with two metallic electrodes available in it. The radioactive emission passing through the volume causes a partial ionization of the gas molecules and generates charged particles in it. The presence of a potential difference between the metallic electrodes causes the current running through the gas volume whose current carriers are the ionization products.

The objective of this article is the obtaining of current-voltage characteristic of a parallel-plane ionization chamber.

This problem is not a new one. The ionization chamber is the device by means of which the radioactivity is investigated [1] up to the discovery of the Geiger-Müller counter. Attempts for the analytical describing of current-voltage characteristic have been done in the right direction [1, 2], but a solution is not found up to now [3]. Besides, the achievements in the experiment and the theory of the physicochemical processes in plasma during last decades [4] provide a good basis for re-examining the solutions of some old problems from the point of view of the modern concepts.

## 2. Elementary processes and balances of particles

During the currency of radioactive emission **R** through the gas volume an ionization process of the gas molecules $A_2$ takes place according the reaction [1, 2, 4]:

$$A_2 + R \xrightarrow{\nu_i} A_2^+ + e + R \tag{1}$$

In the process of ionization positive gas ions and electrons are uniformly generated in the volume possessing the following rate

$$\left(\frac{\delta n_+}{\delta t}\right)_{ion} = \left(\frac{\delta n_e}{\delta t}\right)_{ion} = +\nu_i \cdot N = I_i \tag{2}$$

where $n_+$ is the concentration of the positive gas ions;

$n_e$ is the concentration of electrons in the gas volume;

$\nu_i$ is the frequency of ionization according to reaction (1);

$N$ is the concentration of the neutral gas molecules;

$I_i$ is the ionization rate per unit of volume.

The process (1) is the mechanism through which charged particles are generated in the volume. Thus these charged particles could be lost mainly by two mechanisms:

- Neutralization of the metallic electrodes, when the charged particles drop onto their surface. At low voltages between the metallic electrodes the charged particles clashes with the metallic electrodes do not cause an emission of new electrons from them. The neutralization of the charged particles upon the metallic electrodes is the mechanism for the current closure of the external circuit through the gas volume.
- Volume recombination. During the interaction between a positive ion and an electron in the gas volume they can be neutralized according the reaction [1, 2, 4]:

$$A_2^+ + e \xrightarrow{\beta} A_2 \tag{3}$$

At volume recombination the charged particles are neutralized with the rate

$$\left(\frac{\delta n_+}{\delta t}\right)_{rec} = \left(\frac{\delta n_e}{\delta t}\right)_{rec} = -\beta \cdot n_+ \cdot n_e \tag{4}$$

where $\beta$ is the coefficient of two-particle recombination.

When a potential difference is applied between the metallic electrodes an electric field appears inside the volume which causes the appearance of directed flows of charged particles, i.e. current with densities as following:

$$\vec{j}_e = -e.n_e.\vec{V}_{de} = -e.n_e.\mu_e.\vec{E}, \tag{5}$$

$$\vec{j}_+ = e.n_+.\vec{V}_{d+} = e.n_+.\mu_+.\vec{E}, \tag{6}$$

where $\vec{V}_{de}$ is the drift velocity of electrons in the gas;

$\vec{V}_{d+}$ is the drift velocity of positive ions in the gas;

$\mu_e$ is the mobility of electrons in the gas;

$\mu_+$ is the mobility of positive ions in the gas.

Here we assume that there are small gradients of electrons and positive ions concentrations in the volume, wherefore the diffusion flows are insignificant [4].

Then for the balance of charged particles concentrations (a law for conservation the particles number) we write [4]:

$$\frac{\partial n_e}{\partial t} + \nabla.\left(\frac{\vec{j}_e}{-e}\right) = I_i - \beta.n_e.n_+, \tag{7}$$

$$\frac{\partial n_+}{\partial t} + \nabla.\left(\frac{\vec{j}_+}{e}\right) = I_i - \beta.n_e.n_+. \tag{8}$$

In respect to the density of the electric charge **k** and its current $\vec{j}$ the following relations are valid:

$$k = -e.n_e + e.n_+, \tag{9}$$

$$\vec{j} = \vec{j}_e + \vec{j}_+, \tag{10}$$

$$\frac{\partial k}{\partial t} + \nabla.\vec{j} = 0. \tag{11}$$

The last equation is the law for conservation of electric charge.

In the stationary case the partial derivatives of time in the balances are zeroes.

These are the basic elementary processes with the participation of the charged particles and the concentrations balances of the particles and the electric charge. To this system of equations it is necessary to be added Gauss' law of electric-field strength $\vec{E}$:

$$\nabla \cdot \vec{E} = \frac{k}{\varepsilon_0} = \frac{-e.n_e + e.n_+}{\varepsilon_0}. \tag{12}$$

When considering the ionization chamber irradiated with relatively weak radioactive emission the frequency of ionization is low and the concentrations of the charged particles are small, and they will not change the electric field created by both metallic electrodes, which means that the right side of equation (12) will be nullified [4].

## 3. A plane case
### 3.1. *Geometry and equations*
We will consider farther on a plane case of ionization chamber. Thus both metallic electrodes account for parallel metallic planes situated by distance **d** apart.

The plates have area **S** (**d$^2$**<<**S**). We assume that the electric field is concentrated merely in the volume between the plates and that it is homogeneous.

For describing the spatial coordinates we choose a coordinate system with **OX**-axis perpendicular to the plates and **OY**- and **OZ**-axes forming a plane parallel to the plates.

Let the cathode have coordinate $x_k=0$ and the anode have coordinate $x_a=d$. At such configuration of the electrodes the vectors of the electric-field strength and the motion rates of charged particles will be directed parallel to **OX**-axis. The vector of electric-field strength is pointed from the anode to the cathode. The electrons move in the direction from the cathode to the anode, and they are absorbed by the anode. The positive ions move in the direction from the anode to the cathode, and they are neutralized by the cathode.

If we express from equations (5) and (6) the concentrations of the charged particles by their currents and mobilities, from the balance of charged particles concentration we obtain [4]:

$$\frac{dj_e}{dx} = e.I_i - e.\beta \cdot \frac{j_e}{e.\mu_e.E} \cdot \frac{j_+}{e.\mu_+.E}, \tag{13}$$

$$\frac{dj_+}{dx} = -e.I_i + e.\beta \cdot \frac{j_e}{e.\mu_e.E} \cdot \frac{j_+}{e.\mu_+.E}, \tag{14}$$

$$\frac{dj}{dx} = \frac{d(j_e + j_+)}{dx} = 0. \tag{15}$$

From equation (15) follows that in the plane case **j** is constant in the gas volume. Whereas the densities of current of electrons (5) and ions (6) could be changed in the gas volume, their sum (10) is compulsory constant.

As a first step it is necessary the solutions of equations (13), (14) and (15) considered as a system to be found. We are looking for a solution satisfying the following boundary conditions upon the electrodes [4]:

- cathode does not emit electrons:

$$j_e(x = 0) = 0 \tag{16}$$

- cathode neutralizes the positive ions falling upon it:

$$j_+(x = 0) = j \tag{17}$$

- anode absorbs the electrons falling upon it:

$$j_e(x = d) = j \tag{18}$$

- anode does not emit positive ions:

$$j_+(x = d) = 0 \tag{19}$$

We have chosen to find firstly the function $j_e(x)$. Expressing the current density of the positive ions through the current density of the electrons we obtain:

$$\frac{dj_e}{dx} = e.I_i - \frac{\beta}{e.\mu_e.\mu_+.E^2} \cdot j_e.(j - j_e) \tag{20}$$

Thus we have got an equation that will be mainly studied farther on.

### 3.2. *Regime: No recombination*

We consider this case because it will allow us to study the extreme case when either there is no recombination or the combination is slightly small, e.g. at very great electric field.

Actually, in this case in the right side of equation (20) remains only the first term which does not depend on **x**.

The solution is simple:

$$j_e(x) = e.I_i.x = e.I_i.d.\frac{x}{d} = j_s.\frac{x}{d} \tag{21}$$

Whence using equation (18) we will obtain

$$j = e.I_i.d = j_s. \tag{22}$$

The function $j_e(x)$ (21) is a solution of equation (20) such that the right side of equation (20) has maximum magnitude (if a recombination is available the right side has a smaller value than that).

Translated into physics language this means that all generated charges in the chamber volume are directed to the electrodes and they reach them. That is the reason the current that is available at this regime to be maximum, and we will call it a *current of saturation* $j_s$.

### 3.3 *Regime: With recombination*
When solving the equation (20) in the common case in order to simplify the notations we make it dimensionless to $j_e$ and **x**.

$$z = \frac{x}{d} \in [0,1], \tag{23}$$

$$f(z) = \frac{j_e}{j} \in [0,1]. \tag{24}$$

The function **f(z)** shows how the electron component of current in the chamber volume changes during certain regime of working. Besides, **1-f(z)** will give us an information how the ionic component of current in the volume will change.

Using (23) and (24) equation (20) could be transformed in the following

$$\frac{df}{dz} = \frac{e.I_i.d}{j} - \frac{\beta.d.j}{e.\mu_e.\mu_+.E^2}.f.(1-f) \tag{25}$$

Besides, introducing the substitution

$$4.E_1^2 = \frac{\beta.d.j_s}{e.\mu_e.\mu_+} \tag{26}$$

and using (22) we can carry out (25) in the form

$$\frac{df}{dz} = \frac{j_s}{j} - \frac{4.E_1^2}{E^2} \cdot \frac{j}{j_s} . f.(1-f). \tag{27}$$

This equation (27) should be set in a standard form. For the purpose we get

$$\frac{df}{dz} = \frac{j_s}{j} - \frac{E_1^2}{E^2}\frac{j}{j_s} + \frac{4.E_1^2}{E^2}\frac{j}{j_s}\left(f - \frac{1}{2}\right)^2 = a_0 + a_2\cdot\left(f - \frac{1}{2}\right)^2 \tag{28}$$

where
$$a_0 = \frac{j_s}{j} - \frac{E_1^2}{E^2}\frac{j}{j_s}, \tag{29}$$

and

$$a_2 = \frac{4.E_1^2}{E^2}\cdot\frac{j}{j_s}. \tag{30}$$

According [5] the solution of (28) gives the following

$$\frac{1}{\sqrt{a_0.a_2}}\cdot\arctan\left[\sqrt{\frac{a_2}{a_0}}\cdot\left(f - \frac{1}{2}\right)\right] = z - C. \tag{31}$$

After taking into account the boundary conditions over cathode and anode, and after certain transformations, is obtained

$$f(z) = \frac{1}{2} + \sqrt{\frac{a_0}{a_2}}\cdot\tan\left[\sqrt{a_0.a_2}\cdot\left(z - \frac{1}{2}\right)\right] \tag{32}$$

In order to satisfy the boundary conditions is necessary and sufficient

$$\sqrt{\frac{a_0}{a_2}}\cdot\tan\left[\sqrt{a_0.a_2}\cdot\left(\frac{1}{2}\right)\right] = \frac{1}{2} \tag{33}$$

This equation represents one transcendent equation for the relation between $a_0$ and $a_2$, but in this case it plays the role of an analytical form of the ionization chamber current-voltage characteristic.

**4. Analysis of the solution**

Using solution (32) we present the right side of equation (27) graphically in figure 1. As it is evident (see figure 1.) a great dynamic change of the electron current particularly close to the electrodes is available. Really, close to the electrodes one of the current components is nullified and because of this the lost of charged particles through recombination is small there. In figure 1. is also seen that at weak electric fields the recombination consumes very strongly the generated charged particles inside the volume. This indicates that the differential equation (20) is written in a correct form.

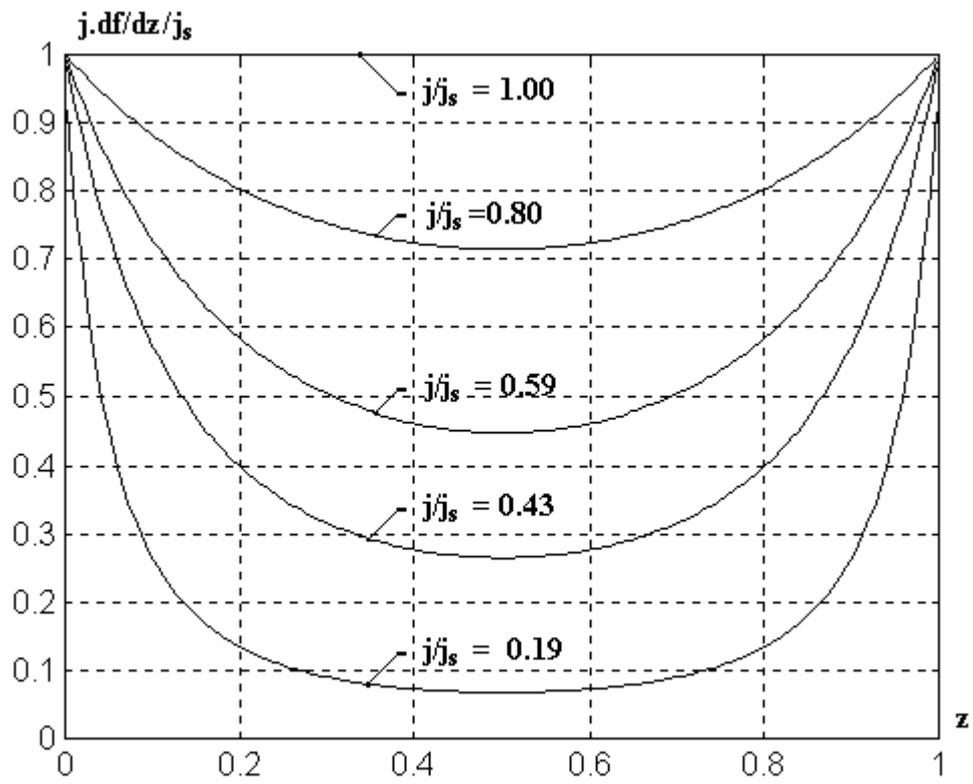

**Figure.1. The right side of equation (27) in relative units.**

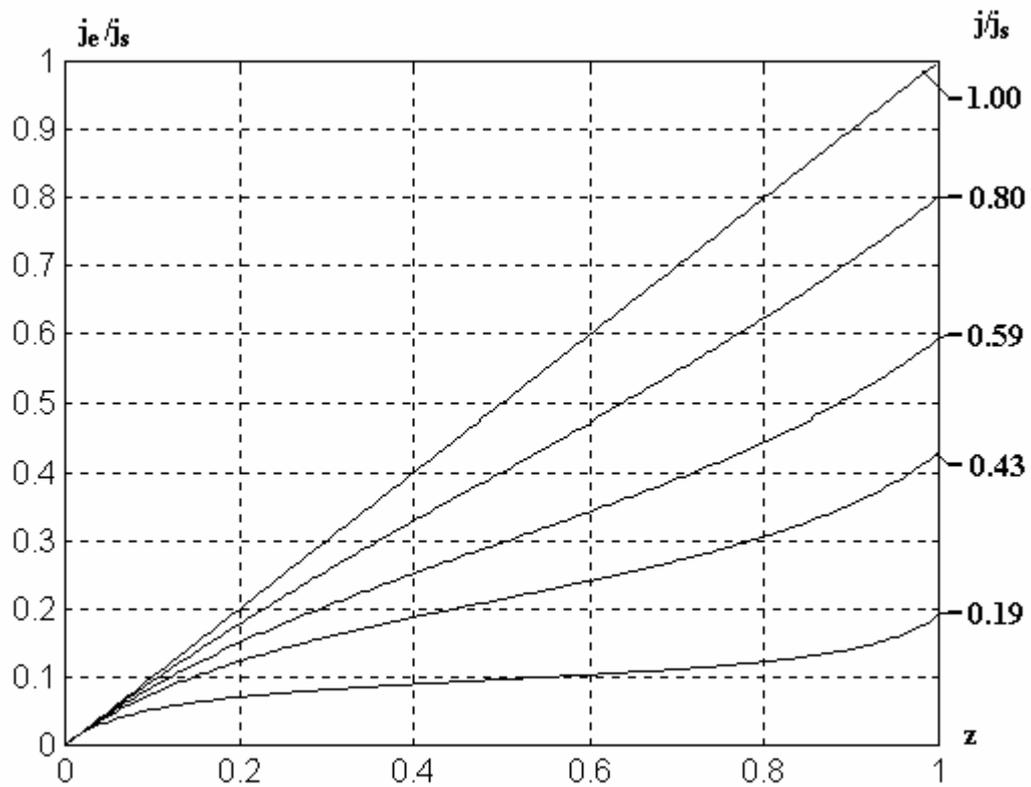

**Figure 2. A curve of set of solutions of (32) at different rate of recombination.**

As a consequence of it the obtained solution (32) has the following graphical form (figure 2).

Finally equation (33) contains the current-voltage characteristic of the parallel-plane ionization chamber. Replacing the constants by their equals from (29) and (30), and making some transformations, we abtain

$$\sqrt{\dfrac{\dfrac{E^2}{E_1^2}-\dfrac{j^2}{j_s^2}}{\dfrac{j^2}{j_s^2}}}\cdot\tan\left[\dfrac{\sqrt{\dfrac{E^2}{E_1^2}-\dfrac{j^2}{j_s^2}}}{\dfrac{E^2}{E_1^2}}\right]=1 \qquad (34)$$

Equation (34) is transcendent and could be solved numerically. The graphical dependence between the current density and the electric-field strength (the potential difference between both electrodes, respectively) is represented in figure 3.

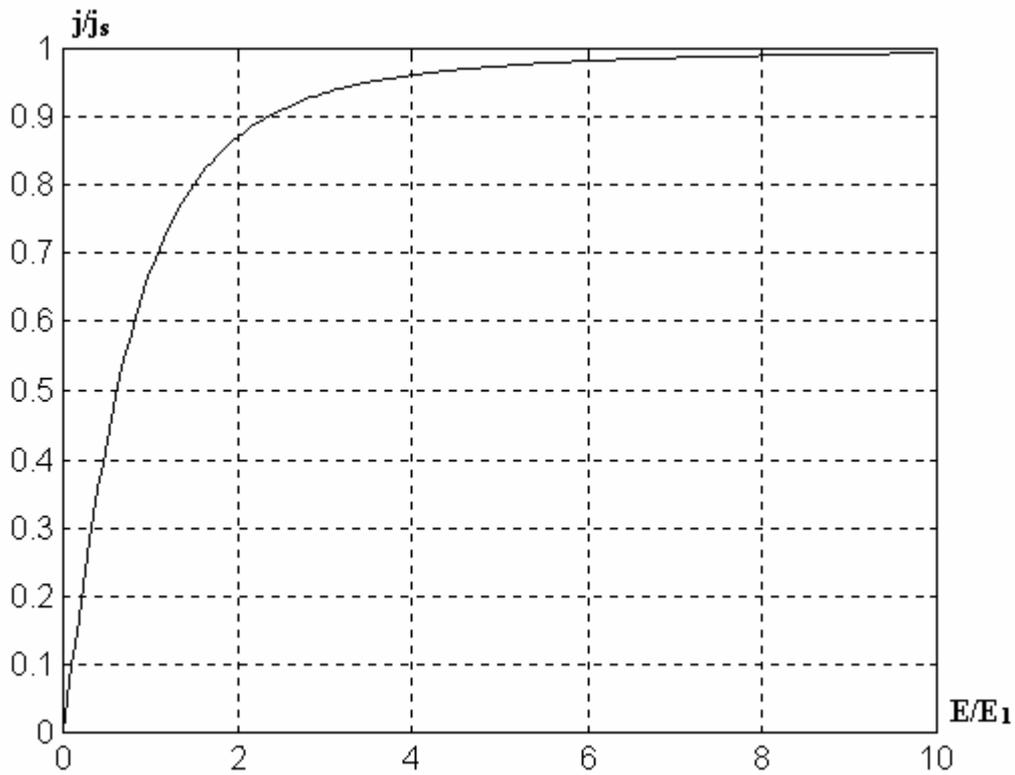

**Figure 3. A current-voltage characteristic of plane ionization chamber**

In table 1 some values satisfying (34) could be read.

Table 1. A current-voltage characteristic of ionization chamber.

| $\dfrac{E}{E_1}$ | $\dfrac{j}{j_s}$ |
|---|---|
| 0.00 | 0.0000 |
| 0.10 | 0.0990 |
| 0.20 | 0.1931 |
| 0.30 | 0.2798 |
| 0.40 | 0.3584 |
| 0.50 | 0.4286 |
| 0.60 | 0.4909 |
| 0.80 | 0.5941 |
| 1.00 | 0.6763 |
| 1.20 | 0.7347 |
| 1.40 | 0.7818 |
| 1.60 | 0.8185 |
| 1.80 | 0.8474 |
| 2.00 | 0.8703 |
| 2.50 | 0.9100 |
| 3.00 | 0.9344 |
| 4.00 | 0.9612 |
| 6.00 | 0.9821 |
| 8.00 | 0.9898 |
| 10.00 | 0.9934 |

As it is evident (figure 3.) with the increase of the electric-field strength the current through the ionization chamber increases as it trends towards the current of saturation. At strong electric fields the current-voltage characteristic could be approximately defined from (34) as:

$$\frac{j}{j_s} \cong 1 - \frac{2}{3} \cdot \left(\frac{E_1}{E}\right)^2 + \frac{4}{5} \cdot \left(\frac{E_1}{E}\right)^4 \qquad (35)$$

When the electric fields are weak the recombination consumes strongly the charged particles generated in the volume, and that is the reason the current to be small. In this case the current-voltage characteristic could be approximately defined from (34) as:

$$\frac{j}{j_s} \cong \frac{E}{E_1} \cdot \left[ 1 - \frac{\pi^2}{8} \cdot \left( \frac{E}{E_1} \right)^2 \right] \tag{36}$$

### 5. Comparison with the experiment

In order to be made a comparison with the experiment is necessary the published yet articles containing data for a large range of currents (from 0 to 100 % of the current of saturation) through a parallel-plane ionization chamber and applied voltages to be chosen. These data should be with a great precision in the whole range of conditions.

Therefore the reference [6] was noticed in which the creation of a parallel-plane ionization chamber (with $d = 0.3 \, cm$), allocated for medical irradiations, is reported. In this work the results from the measurement of the current $i$ through the ionization chamber are given for series of four air pressures in the chamber volume and for four intensities of ions beam $C^{6+} (290 MeV/u)$, as for each series the measurement are done for at least 15 magnitudes of voltages applied to the chamber. Especially useful for the measurements precision in [6] is turned out the application of the current $i_{cont}$ of another ionization chamber for the control of beam intensity and for the correction of the obtained results according this current. In this case the control ionization chamber is working in a regime of saturation and neither the gas pressure nor the applied voltage to it is changing.

Here we will use only the results from the series with a change in the working pressure of the gas in the chamber at one and the same intensity of ions beam. For these series the dependence between the relative effectiveness $f = i/i_{cont}$ of the investigated ionization chamber in comparison with the control one and the applied voltage $U$ is represented graphically in [6].

We suppose that at a constant intensity of the ionizing irradiation

$$\frac{j}{j_s} = \frac{f}{f_{sat}}, \tag{37}$$

and

$$\frac{E}{E_1} = \frac{U}{U_1}. \tag{38}$$

**Table 2. Values of $f_{sat}$ and $U_1$ at different pressures of the gas in the chamber volume**

| P, Torr | $f_{sat}$ | $U_1$, V |
|---|---|---|
| 759.6 | 0.780 | 151.0 |
| 202.2 | 0.213 | 23.3 |
| 111.2 | 0.109 | 12.1 |
| 50.2 | 0.049 | 5.3 |

The new magnitudes $f_{sat}$ and $U_1$ are constants within the framework of one series (for one and the same gas pressure) but their values are various for the different series. In this work they are obtained as a result of a selection and a fitting.

Using (37), (38) and the data represented in Table 2 the data from [6] are shown in Figure 4.

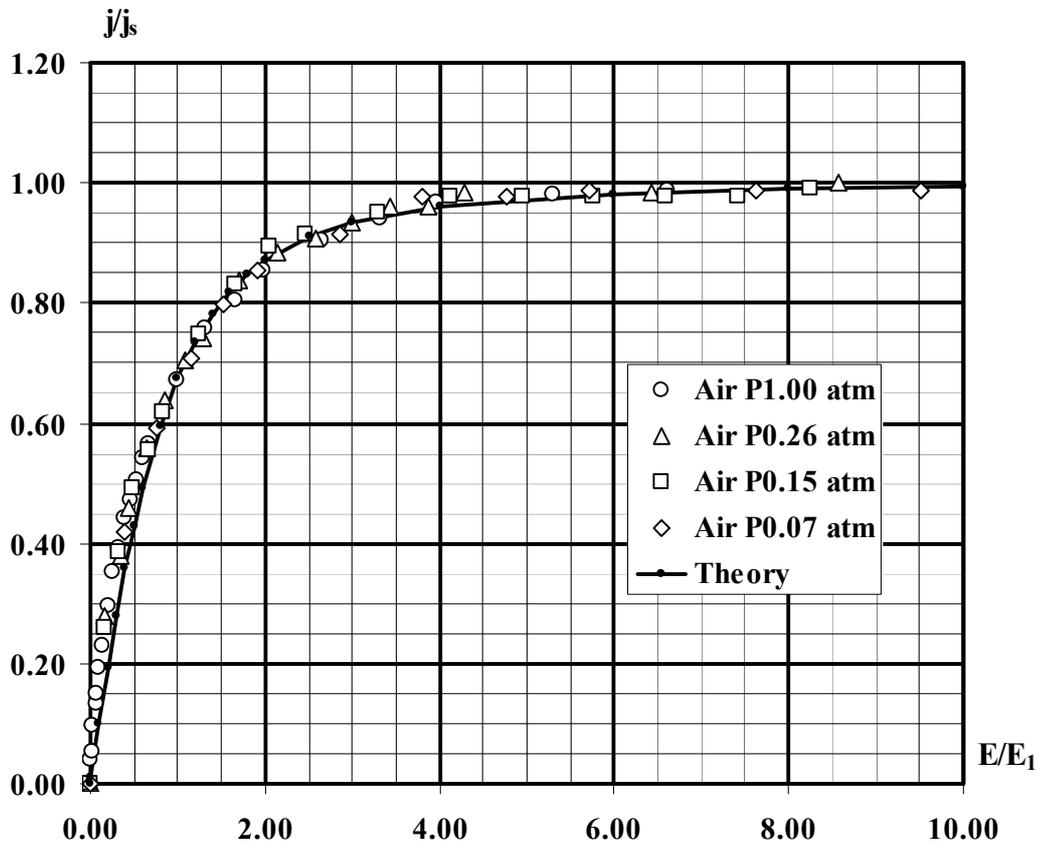

**Figure 4. A comparison between the theoretical curve and the experimental points from [6].**

As it is evident from Figure 4, in the investigated range of values of the electric field strength, the theoretical curve moves either close or between the experimental results. Therefore we may say that as a dynamics of the dependence course and as an agreement between the values there is a good accordance between the theoretical curve and the experimental results.

## 6. Conclusion

In conclusion it could be said that a differential equation is derived which the current submits to in the volume of a plane ionization chamber. An analytical formula of the current-voltage characteristic of a parallel-plane ionization chamber is obtained in the form of a transcendent equation.